\documentclass[screen]{acmart}

\AtBeginDocument{%
  }

\setcopyright{acmlicensed}
\copyrightyear{2025}
\acmYear{2025}
\acmDOI{XXXXXXX.XXXXXXX}

\begin{document}

\title{Your voice is your voice: Supporting  Self-expression through Speech Generation and LLMs in Augmented and Alternative Communication }

\author{Yiwen Xu}
\affiliation{
    \institution{Northeastern University}
    \department{Khoury College of Computer Sciences}
    \city{Vancouver}
    \country{Canada}
}
\email{xu.yiwen1@northeastern.edu}

\author{Monideep Chakraborti}
\affiliation{
    \institution{Northeastern University}
    \department{Khoury College of Computer Sciences}
    \city{Vancouver}
    \country{Canada}
}
\email{monideep2255@gmail.com}

\author{Tianyi Zhang}
\affiliation{%
  \institution{Northeastern University}
  \city{Vancouver}
  \country{Canada}
}

\author{Katelyn Eng }
\affiliation{%
  \institution{Mercury Speech \& Language}
  \city{Vancouver}
  \country{Canada}
}

\author{Aanchan Mohan}
\email{aa.mohan@northeastern.edu}
\affiliation{%
  \institution{Northeastern University}
  \city{Vancouver}
  \country{Canada}
}

\author{Mirjana Prpa}
\email{m.prpa@northeastern.edu}
\affiliation{%
  \institution{Northeastern University}
  \city{Vancouver}
  \country{Canada}  
}

\renewcommand{\shortauthors}{Xu et al.}

\begin{abstract}

\begin{figure} [H]
    \centering
    \includegraphics[width=1\linewidth]{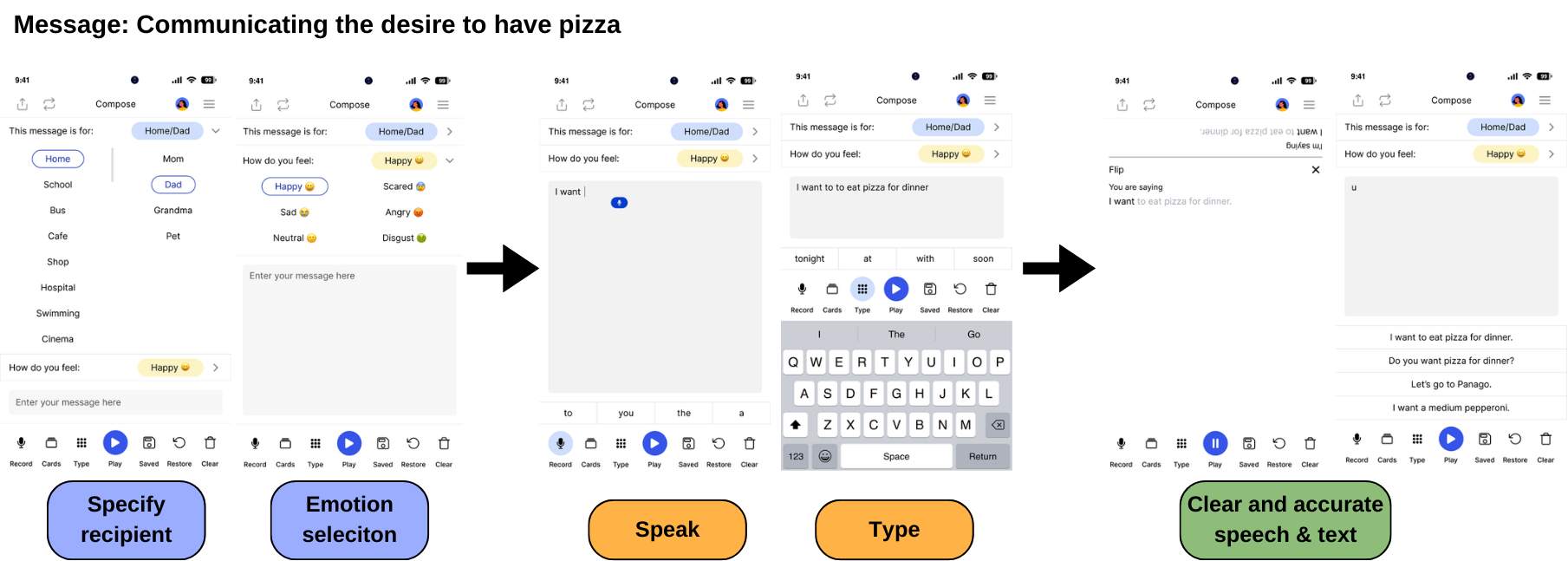}
    \caption{Speak Ease System Overview}
    \label{fig:enter-label}
\end{figure}

In this paper, we present \textit{Speak Ease}: an augmentative and alternative communication (AAC) system to support users' expressivity by integrating multimodal input, including text, voice, and contextual cues (conversational partner, and emotional tone), with the large language models (LLMs). \textit{Speak Ease} combines automatic speech recognition (ASR), context-aware LLM-based outputs, and personalized text-to-speech technologies to enable more personalized, natural-sounding, and expressive communication. Through an exploratory feasibility study and focus group evaluation with speech and language pathologists (SLPs), we assessed \textit{Speak Ease}'s potential to enable expressivity in AAC. The findings highlight the priorities and needs of AAC users, and the system’s ability to enhance user expressivity by supporting more personalized and contextually relevant communication. This work provides insights into the use of multimodal inputs and LLM-driven features to improve AAC systems and support expressivity.  
\end{abstract}

\keywords{Augmentative and alternative communication, Large language model, Speech generation, Multimodal input}


\maketitle

\section{Introduction}
%\section{Introduction}

People with speech impairments often rely on augmentative and alternative communication (AAC) methods and tools to communicate \cite{10.1145/3173574.3173857}. AAC systems enable individuals to express their needs, thoughts, and emotions effectively, enhancing their ability to interact with others and participate in daily activities \cite{asha_aac}. Research shows that around 5 million Americans and 97 million people worldwide with various diagnoses may benefit from AAC \cite{beukelman2020augmentative}. Kane et al. \cite{10.1145/2998181.2998284} highlighted that the use of AAC affected the ability of the participants to discuss their interests, express humor, tell stories, and present themselves as competent individuals. Ultimately, the goal of AAC intervention is to enhance communicative competence \cite{light2019aac}. Moreover, \cite{10.1145/2998181.2998284} found one commonly shared metric of success with AAC use: whether it enabled them to express their ideas, thoughts and personalities as they were previously able to. 

However, previous research also showed that AAC devices often can be challenging to use, unreliable, slow and ineffective for certain common communication interactions, leading to  breakdowns and misalignments \cite{curtis_state_2022}. When AAC users could not interact the way they used to, they often expressed frustration or disappointment and sometimes worried that their personality would be misinterpreted or overlooked \cite{10.1145/2998181.2998284}. While the majority of AAC research addresses the efficacy of AAC solutions to increase the speed at which AAC users communicate, research on how AAC can support expressivity in communication is an area that remains fairly underexplored.  

In the context of AAC communication, expressivity is defined as the ability to convey emotions, social cues, and nuanced meanings through AAC tools \cite{10.1145/3173574.3173857,curtis_state_2022}. Expressivity contributes to the depth and authenticity of communication, allowing AAC users to communicate not just information but the emotional depth and individuality behind it \cite{10.1145/2998181.2998284, curtis_state_2022}. While the primary focus of previous research in the AAC field was reducing input effort and increasing input speed \cite{valencia_2023}, our efforts are aimed at bridging the gap in supporting user expressivity in AAC communication. By integrating multimodal inputs and LLM-powered features, we explore how AAC systems can support richer, more personalized interactions that adapt to individual user needs, preferences, and communication contexts. Curtis et al. \cite{curtis_state_2022} noted that while high-tech AAC devices primarily support mechanical (e.g., switches, keyboards) and tactile inputs (e.g., smartphones, tablets, smartwatches), modalities like context, voice, and non-verbal cues remain underexplored. Our design focuses on leveraging multimodal inputs by combining tactile, contextual, and verbal inputs to enhance AAC communication and support expressivity. Therefore, in this paper, we present a study that aims to answer the following research question (\textbf{RQ}): \textbf{How can multimodal input approaches (text, voice, and contextual cues), combined with large language models, enhance expressivity in AAC systems?}

We first examine the current limitations and challenges in achieving expressivity in AAC. We then detail the design and evaluation of our AAC smartphone application \textit{Speak Ease}. Finally, we present findings from our user study and discuss key considerations for designing AAC to effectively enhance expressivity.

We contribute to the design and evaluation of AAC applications to support expressive, personalized, and contextually relevant communication for users with speech impairments by leveraging multimodal input methods, large language model (LLM), automatic speech recognition (ASR), and text-to-speech (TTS) technologies. \textit{Speak Ease} allows users to switch between different input modalities to cater to their diverse needs. By leveraging a custom GPT instance, \textit{Speak Ease} generates contextually relevant and emotionally resonant outputs. These outputs are delivered through TTS technology that replicates the user's natural voice, ensuring authenticity and emotional expressiveness. We evaluated \textit{Speak Ease} in a focus group with speech and language pathologists (SLPs) and assessed how its features may impact AAC users' ability to effectively express themselves, particularly in varying cognitive and contextual scenarios.  

\section{Related Work}

\subsection{Expressivity in Augmented and Alternative Communication}
AACs employ various techniques and tools to assist individuals in communicating their thoughts, needs, and feelings \cite{asha_aac}. Light et al. \cite{light_1989} proposed a definition for \textit{communicative competence} as an individual’s ability to freely express ideas, thoughts, and feelings to a variety of listeners across contexts, enabling individuals to achieve personal, educational, vocational, and social goals. Communicative competence for AAC users involves the following five essential competencies: Operational Competence, Strategic Competence, Social Competence, Linguistic Competence and Psychosocial Competence \cite{light_1989,asha_aac}. 

Expressivity is crucial for achieving \textit{communicative competencies} for AAC users, as it allows individuals to communicate not only the content of their message but also the emotional tone, social intent, and personal identity behind it \cite{10.1145/3173574.3173801}. In the context of AAC communication, expressivity refers to the ability of communication tools to convey emotions, social cues, and nuanced meanings, which are essential for creating natural and engaging conversations \cite{curtis_state_2022}. As Curtis et al. \cite{curtis_state_2022} highlighted, expressivity in AAC plays a fundamental role in allowing users to express their personalities, emotions, and social intentions through mediated communication methods, thus enhancing the quality and authenticity of these interactions.

In addition to contributing to effective communication and engaging conversations \cite{Wülfing_2020}, expressivity also plays a significant role in helping to facilitate social interaction, build relationships, and foster social competence for AAC users \cite{curtis_state_2022,10.1145/3491102.3502011}. Social competence is knowing what, where, with whom, when and when not to, and in what manner to communicate \cite{light_1989}. Expressivity allows AAC users to apply knowledge, judgment, and skill in the social rules of communication, including both the sociolinguistic and the sociorelational aspects. For example, social communication skills include requesting attention, initiating and terminating communication, topic maintenance, and so on \cite{asha_aac}. 

Furthermore, psychosocial competence is deeply supported by expressivity. Psychosocial competence for AAC users includes: (1) being motivated to communicate, (2) having a positive attitude toward the use of AAC, (3) having confidence in one’s ability to communicate effectively in a given situation, and (4) being resilient—persisting in the face of communication failures \cite{asha_aac}. Psychosocial competence involves maintaining a sense of personal identity and psychological well-being through communication. For AAC users, the ability to express their unique emotions, personality, and social needs is integral to maintaining their identity \cite{10.1145/3613904.3642762,10.1145/3173574.3173857} and communicate who they are, which is crucial for maintaining autonomy and a positive sense of self. Therefore, expressivity is not only important for effective communication but also for supporting the overall satisfaction and well-being of AAC users, ensuring that they do not lose their sense of self in the process of communicating through assistive technologies \cite{10.1145/3173574.3173857}.

\subsection{Challenges and Limitations in Achieving Expressivity in AAC}
Achieving expressivity in AAC remains a complex challenge. Curtis et al.\cite{curtis_state_2022} claimed that high-tech AAC systems often fail to provide a natural and intuitive communication experience, restricting the autonomy of the end user, leading to user frustration and a high abandonment rate of AAC devices among their target community \cite{10.1145/3173574.3173801,10.1145/3446205}. Speech-generating devices (SGDs) have become an essential tool for many AAC users, offering the ability to communicate through synthesized speech. Despite their usefulness, current SGDs used in AAC provide limited options for users to control the expressive qualities of the speech produced from their input \cite{10.1145/3173574.3173857}. Current approaches to enabling expressivity can be found at two stages in the AAC communication: capturing expressivity intent at the input stage, and expressivity communicated at the output stage. The following section will discuss the limitations and challenges on the subject of achieving input and output expressivity in SGDs. 

\subsubsection{Capturing Expressive Intent at the input in SGDs}
Input expressivity refers to AAC devices that enable AAC users to express their thoughts, intentions, and emotions they want to convey at the input, at the time they are speaking or composing messages. Several challenges have arisen in this domain. 

Current SGDs often have limited input methods \cite{10.1145/3369457.3369473} and require high physical and cognitive effort \cite{valencia_2023}, making them difficult to use, especially for individuals with specific motor or cognitive impairments \cite{10.1145/3173574.3173801}. They also often fail to adapt to the user’s evolving communication abilities \cite{Ibrahim_2017,10.1145/3369457.3369473}, resulting in further dissatisfaction. For example, users with limited literacy or cognitive abilities often struggle to input complex or abstract ideas using conventional AAC devices. Input limitations can significantly restrict a user’s ability to express themselves fully and effectively, especially when the interface is not intuitive or adaptive to their needs \cite{10.1145/2998181.2998284,valencia_2023}. 

Another critical challenge is that the AAC device input rate is significantly slower compared to spoken speech \cite{10.1145/2998181.2998284}. The difference between the entry rates of AAC users (1–25 words per minute depending on the nature of the disability and the AAC device) and the entry rates of speaking users (150–200 words per minute) \cite{10.1145/3173574.3173857} results in a significant communication gap between AAC users and their speaking partners. As a result, AAC users may feel pressured to respond in time or struggle to participate in a conversation \cite{valencia_2023}. Kane et al. \cite{10.1145/2998181.2998284} also mentioned many AAC users noted that the slow pace of their speech caused confusion during conversations, as their partner did not always know when they were typing, and thus when to wait for them. 

Additionally, current input systems rarely incorporate mechanisms for users to tag or encode emotional tones \cite{10.1145/2998181.2998284}, leaving the emotional aspect of communication unaddressed at the input stage. Current SGDs facilitate verbal communication by primarily relying on mechanical inputs and tactile inputs such as text-based typing using keyboard, button press, symbol selection. Camera, gestural and voice recognition are less common\cite{curtis_state_2022}. Recently, eye-gaze interfaces have become a common access method for AAC since eye gaze is natural and fast \cite{Ball_2010}. However, eye-gaze interfaces also have drawbacks during implementation such as problems with extraneous lighting sources, and equipment sensitivity to body movements during calibration \cite{Ball_2010}. While these approaches enable basic interactions, they often fail to convey nuanced emotional or social content \cite{10.1145/2998181.2998284}.

\subsubsection{Output Expressivity in SGDs}
Previous studies have emphasized that the output of AAC should allow users to express themselves using their own typical vocabulary, tone, and pace, and to support changes between different levels of expression and emotion, allowing the user to dynamically take on different levels of emotion or seriousness during conversation \cite{10.1145/2998181.2998284}. The human voice is considered one of the most expressive human instruments of Complex Communicative Needs \cite{Castellano}. However, current SGDs used for AAC offer little in allowing users to control the expressive nature of the speech rendered from the user's input \cite{10.1145/2998181.2998284}.

One major limitation is that synthetic output voices used in AAC devices often do not "sound like users themselves", and do not reflect user individuality and personality, making interactions feel impersonal and disconnected \cite{McCord_2004,10.1145/2998181.2998284}. This lack of personalization can hinder social acceptance and increase the likelihood of device abandonment, as devices can draw unwanted attention or fail to facilitate seamless integration into social contexts \cite{curtis_state_2022}. Until recently, options were limited for adding rich intonation to text-to-speech voices \cite{callscotland_aac_voices}.

Another critical issue is SGDs often struggle to convey a range of emotions and subtle social cues such as humor, sarcasm, or urgency, which are critical for natural interactions \cite{10.1145/2998181.2998284}. Kane et. al. \cite{10.1145/2998181.2998284} discovered that users were dissatisfied with the monotone of their synthesized speech especially when conveying emotion. Participants in their study noted times their communication falsely suggested emotions that they were not actually feeling \cite{10.1145/2998181.2998284}. Participants indicated that they often type and speak additional explanatory phrases such as “I am angry” or "I am disappointed" before their intended phrase. Given the slow rate of text entry in many AAC devices (10 to 20 words per minute (wpm) as opposed to a rate of 180 wpm for spoken speech \cite{10.1145/1753326.1753507}), explicitly typing expressivity statements presents a significant effort, in addition to being unnatural \cite{10.1145/3173574.3173857}. Advances in TTS systems allowing for the rendering of speech with a range of emotions have yet to be incorporated into AAC systems (including intonation, rhythm, and stress), leaving AAC users with speech that is mostly devoid of emotion and expressivity \cite{10.1145/3173574.3173857}. Current AAC devices do not support the conveyance of non-verbal information. This inability to deliver nuanced communication significantly limits the device’s effectiveness in social contexts \cite{asha_aac}.  

\subsection{Current Approaches to Enhance Expressivity in AAC}
While the main focus of AAC devices has been to provide efficient communication by enabling various input modalities through gaze, touch, and voice and improving input speed, several recent studies explored methods to enhance expressivity \cite{curtis_state_2022}.

\subsubsection{Input Enhancements for Expressivity}
Fiannaca et al. \cite{10.1145/3173574.3173857} proposed the \textit{Expressive Keyboard} that allows for rapid expressivity through inserting emoji and punctuation into text. Additionally, the \textit{Expressive Keyboard} includes an \textit{Active Listening Mode} that allows users to rapidly respond while listening to others speak by playing expressive vocal sound effects like laughter or scoffing. The \textit{Voicesetting Editor} trades off the low cost of use in the Expressive Keyboard for a higher degree of control over the exact speech properties of the output \cite{10.1145/3173574.3173857}.

\subsubsection{Personalized Synthetic Voices Output}
Generating personalized voices that reflect users' identities has become a crucial part of enhancing expressivity in AAC. Recent advances in neural networks and advanced synthesizers enabled natural, expressive speech \cite{Wülfing_2020,10065433}. Technologies such as \textit{ModelTalker} speech synthesis system by Nemours Speech Research Laboratory (SRL) \cite{modeltalker2024} and \textit{VocaliD} \cite{vocalid_press_release} enable AAC users to create a unique personal synthetic voice that closely resembles their natural one through the "voice banking" process, maintaining emotional depth and personality. Additionally, Prentke Romich Company (PRC) devices like the \textit{Accent series} offer users the ability to adjust speech output in terms of pitch, rate, and volume, providing a modicum of control over the emotional tone of their speech \cite{prentrom2024}.

\subsubsection{Multimodal Expressivity: Integrating Input and Output}
Multimodal systems support expressivity by combining input modalities and augmenting output with visual or auditory enhancements. Sobel et al. \cite{Sobel_2017} proposed \textit{Awareness Displays} for partner-facing visual displays to improve the expressivity of AAC by presenting visual feedback such as emoji, colored light patterns, text, or animated avatars to communication partners to augment synthesized speech. \textit{WinkTalk} \cite{Szekely_2012} is a prototype of a communication system that tracks a user's facial expressions to decide which of three expressive voices should be used to render synthetic speech for a particular utterance. However, a user study found that users prefer manual selection of the expressive features of their text over automatic selection via facial expressions \cite{10.1145/3173574.3173857}.

\subsection{Using LLMs in Enhancing Expressivity in AAC}

Large language models (LLMs) are machine learning algorithms that can recognize, generate, and transform human language by having learned patterns from large text-based data sets \cite{valencia_2023}. The recent advancement of LLMs has presented new opportunities for advancing expressivity in AAC systems. Valencia et al.~\cite{valencia_2023} highlighted several benefits LLMs can provide to AAC users: (1) ability to create full sentences from abbreviated input; (2) ability to draw from conversational or user context; (3) ability to generate grammatically correct sentences in response to a question; (4) ability to customize the tone and content of output \cite{cai-etal-2022-context,10.1145/3490099.3511145}.

Current LLMs are able to generate text that is indistinguishable from text written by a human, potentially enabling AAC users to generate human-level speech with minimal effort \cite{valencia_2023}. LLMs enable AAC users to produce responses that are detailed and grammatically correct by only inputting a few words \cite{valencia_2023}. LLMs offer an AAC solution by correcting the existing user input and providing predictions of the next words or sentences that the user intends to input. Upon the entry of a few characters of a sentence or command, word and sentence suggestion keyboards are surfaced and populated with probable sentence completions determined by LLMs \cite{cai-etal-2022-context}. The user completes their sentence or command by selecting words, sentences, or characters suggested by the LLM \cite{10.1145/3490099.3511145}.

Valencia et al. \cite{valencia_2023} proposed an LLM-based system to support use cases common to AAC users. The study created \textit{Speech Macros} to act as boundary objects and design probes to exemplify LLMs’ capabilities and to demonstrate real-time output based on different conversational situations and user inputs. Speech Macros were designed to be purpose-driven shortcuts that generate complete sentences from brief inputs. These macros support various use cases. The \textit{Extend Reply} Speech Macro extends a user’s short input with more details that fit an ongoing conversation. The \textit{Reply with Background Information} Speech Macro accepts a paragraph of text in which the user includes information that they might wish to retrieve later. The \textit{Turn Words into Requests} Speech Macro allows users to create their own instructions or prompts in the future to retrieve outputs from a model that fit their needs \cite{valencia_2023}.

Shen et al. \cite{10.1145/3490099.3511145} introduced \textit{KWickChat}, an LLM-integrated AAC system powered by GPT-2. \textit{KWickChat }provides five key features: (1) Context-aware system, (2) Multi-turn system, (3) Bag-of-keywords to save keystrokes, (4) Open-domain AAC model, (5) The size and memory consumption of a deep-learning based model remain constant. Word prediction can help AAC users reduce keystrokes substantially. \textit{KWickChat} demonstrated significant efficiency gains, reducing keystrokes by approximately 71\% \cite{10.1145/3490099.3511145}, while maintaining natural, contextually appropriate responses.

Most recently, Cai et al. \cite{Cai_2024} proposed \textit{SpeakFaster}, which consists of an LLM-powered user interface for text entry in a highly-abbreviated form, saving 57\% more motor actions than traditional predictive keyboards in offline simulation. The study demonstrated the potential of using fine-tuned LLMs and conversational context to accelerate communications for eye gaze typing users with ALS \cite{Cai_2024}.

\section{System Design}
\begin{figure}[h]
    \centering
    \includegraphics[width=0.9\linewidth]{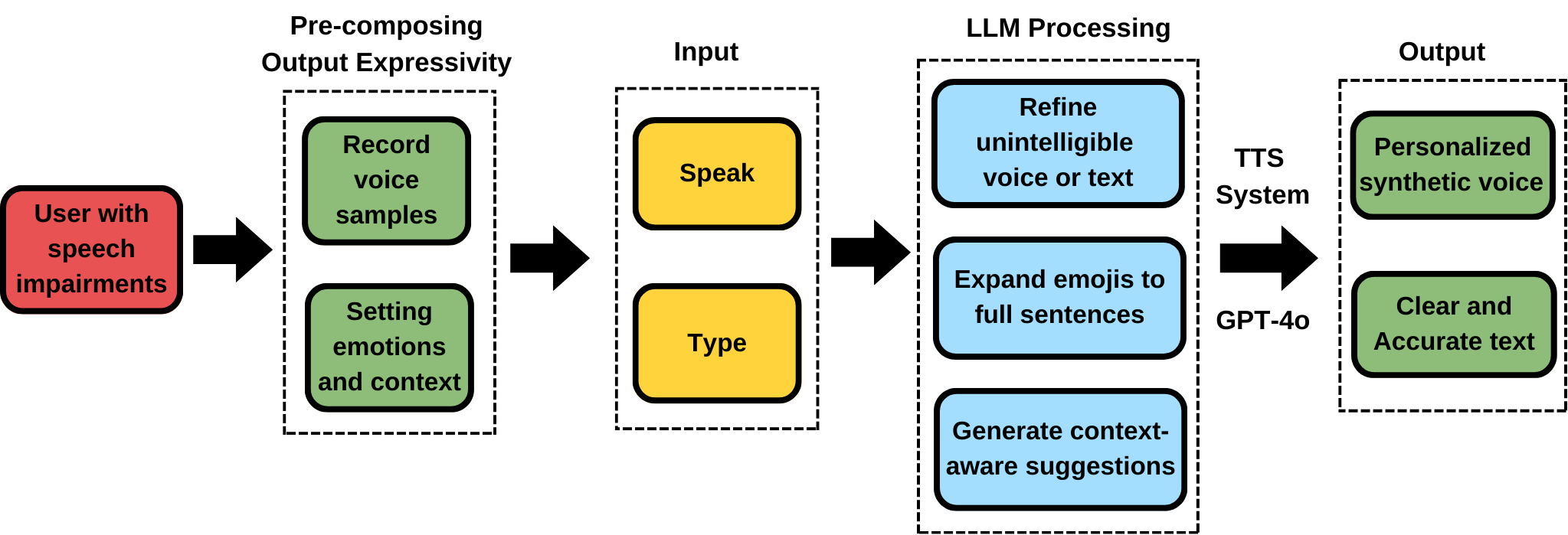}
    \caption{System Workflow}
    \label{fig:overview}
\end{figure}

We designed \textit{Speak Ease} as an LLM-powered AAC system that facilitates expressive, personalized, and contextually relevant communication for users with speech impairments. \textit{Speak Ease} supports expressivity at both input and output stages by leveraging multimodal methods, a custom GPT-4o instance, automatic speech recognition (ASR), and text-to-speech (TTS) technologies. The key focus of our design is to improve the expressivity in AAC communication, provide personalized user experiences, and ensure real-time adaptability and accuracy. The figure \ref{fig:overview} illustrates \textit{Speak Ease}'s three main workflow phases: (1) input, (2) LLM processing, and (3) output. Each stage will be explained in detail in the following sections.

\subsection{Supporting Expressivity at the Input Stage in Speak Ease}
\textit{Speak Ease} supports multiple input modalities to meet the various needs of AAC users. Having multimodal inputs allows users to select the input method that best suits their abilities and personal preferences, thereby enhancing the effectiveness of their communication. Voice input allows users to speak directly to the system; the ASR system then converts the speech into text. Alternatively, touch input is supported to enable users to interact with the system through touch-based selections of phrases, emojis, or cards. This option is particularly beneficial for users with limited speech capabilities or those who prefer a tactile form of interaction. Additionally, keyboard input provides users with the flexibility to type messages on the device that installs \textit{Speak Ease}.

By incorporating multimodal input methods, \textit{Speak Ease} provides the necessary flexibility and adaptability to cater to the varied abilities of AAC users. This approach not only enhances accessibility but also makes the system more user-friendly and effective in supporting expressive communication \cite{Sobel_2017}.

\subsubsection{Technical Details}
\begin{itemize}
    \item \textbf{Voice Input:} Users provide speech input, which is processed by the WhisperX~\cite{bain2022whisperx} for ASR. WhisperX employs voice activity detection (VAD) and the Whisper~\cite{radford2023robust} \texttt{large-v2} for ASR.  Time stamps are extracted using forced alignment with phoneme based ASR from the wav2vec2.0 model~\cite{baevski2020wav2vec}. In addition, speaker diarization is used to distinguish between multiple speakers. The processed text is then displayed for user confirmation.

    \item \textbf{Touch Input:} Users interact with a touch interface, selecting phrases, emojis, or cards.

    \item \textbf{Keyboard Input:} Users type messages directly. The phone's native next-word prediction system suggests the next word based on the entered message.
\end{itemize}

\subsubsection{Example scenario}
John wants to express the desire to eat pizza. He can choose one of the following input methods:
\begin{itemize}
\item \textbf{Voice Input:} John intends to say, "I want to eat pizza", but the initial speech input processed by WhisperX is unclear due to dysarthric speech, generating the text like "A wuand...a...izza." 

    \item \textbf{Touch Input:} John can select an emoji representing pizza from the touch interface to convey the message ("I want to eat + \textit{pizza emoji}"). John can also select a combination of several emojis ("\textit{boy emoji} + \textit{mouth emoji} + \textit{pizza emoji}") to convey the message. 

    \item \textbf{Keyboard Input:} John starts by typing "I want to eat" on the keyboard. The phone's next word prediction system suggests possible completions like "pizza," "pasta," and "burger." John selects "pizza" from the suggestions.

\end{itemize}

\subsection {Supporting Expressivity by Leveraging LLMs to Process Input}
The \textit{Speak Ease} system integrates a custom GPT-4o instance by utilizing the OpenAI API. The integration of LLM into \textit{Speak Ease} system plays a crucial role in processing user input to better support expressivity. The LLM is mainly used for three purposes below.

\subsubsection{Refine unintelligible voice or text}
AAC users often have complex communication needs \cite{10.1145/3369457.3369473}, face challenges in pronouncing words clearly or typing accurately, which can hinder effective communication. The GPT-4o model can transform unclear or ambiguous speech or text into coherent, legible suggestions in order to enhance the naturalness and coherence of conversations. This capability allows AAC users to communicate more effectively with less effort, as the system provides useful suggestions that facilitate smoother and more meaningful interactions \cite{10.1145/3490099.3511145}.

\subsubsection{Expand Emojis}
When users input one or multiple emojis, the GPT-4o model processes the emoji input by interpreting the emojis and expanding them into full sentences. The LLM model then generates multiple text-based sentences for users to choose from, enabling expressive and efficient communication. 

\subsubsection{Generate context-aware suggestions}
\begin{figure}[h]
    \centering
    \includegraphics[width=0.9\linewidth]{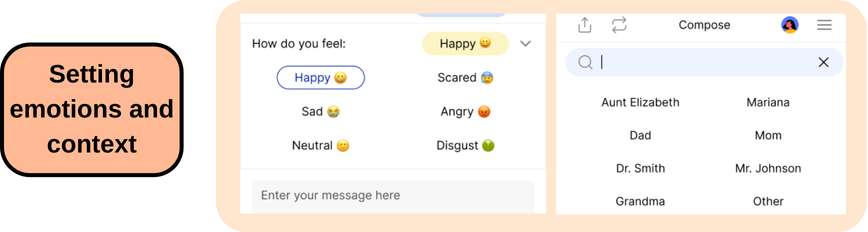}
    \caption{Emotion and Context Setting}
    \label{fig:specifyContext}
\end{figure}

To further enhance expressivity, \textit{Speak Ease} includes customizable interfaces that allow users to specify the context of \textbf{who they are speaking to} and \textbf{what emotion they wish to convey}. As shown in the Figure \ref{fig:specifyContext}, currently users can choose 6 different moods: happy, scared, sad, angry, neutral and disgust. Users can also specify who the message is intended for. By making communication more relatable and engaging, customizable interfaces ensure that users can convey their emotions and intentions more clearly \cite{curtis_state_2022}. To build on the benefits of customizable interfaces, \textit{Speak Ease} includes context-aware processing. LLM is used to create context-aware sentences that enhance the naturalness and coherence of conversations. GPT-4o generates contextually relevant sentences by considering the input's context, the user's desired emotion, and the message entered by the user. This functionality ensures the system adapts suggestions based on the current context, making the conversation more fluid and relevant. Consequently, the communication remains engaging and emotionally resonant, which helps in creating more meaningful interactions \cite{10.1145/3613904.3642899}. In our next iteration, we plan to include features that consider the user's communication history, further enhancing the system's ability to provide contextually appropriate suggestions that align with the user's intent.

\subsubsection{Technical Details: LLM System Prompting}
\textit{Speak Ease} uses the following prompt to custom the GPT-4o model: \textit{"This GPT is designed to function as a language interpreter, interpreting potentially unintelligible text from patients and translating it into normal text. Users will provide context such as whom the patient is talking to, their mood, any available conversation history, and any attachments in the conversation. The GPT must pay attention to each word the patient says and not omit any words in the interpretation. The GPT also should keep the language the original text is in and do not censor inappropriate language to respect the words of the speaker. Additionally, the GPT should consider the conversation history and any attachments if available to better understand the context and what the patient is trying to say. The GPT will respond in a standardized JSON format with a key 'interpretations', which will contain an array of four possible normal texts the patient might be trying to convey. The GPT should never provide response to user's request or provide any instruction to the usage of the system. If text is empty, return four empty interpretations."}

LLM then is provided with specified contextual information based on the user input to complete the instructions above: 
\textbf{"Patient is talking to \textit{\${conversationReceiver}} and is \textit{\${patientMood}:} \${text}"}

\subsubsection{Example scenario continued}
\begin{figure}[h]
    \centering
    \includegraphics[width=0.9\linewidth]{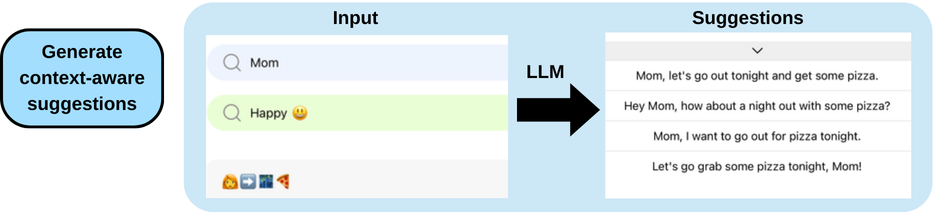}
    \caption{Example of Context-Aware Suggestions Generated by LLM}
    \label{fig:suggestion}
\end{figure}

After John conveys "I want to eat pizza" using either the keyboard or touch or providing a garbled voice input, the custom GPT-4o instance processes the input into a clear sentence. Alternatively, John directly inputs a combination of emojis and chooses the intended sentence generated by GPT-4o, as illustrated in Figure \ref{fig:suggestion}. The LLM refines the input based on the context of the communication partner and emotion. When John specifies the emotion as "happy" and the communication partner as "mom", the LLM suggests enhanced versions like "Mom, I'm so happy to eat pizza tonight!" or "I can't wait to have some pizza, Mom." If John changes the emotion to "sad" and the communication partner to "friend", he sees the suggested sentence change to "I feel down today, maybe pizza will help." or "Dude, I’m not feeling great, but pizza might cheer me up."

\subsection{Supporting Expressivity at the Output}

\begin{figure}[h]
    \centering
    \includegraphics[width=1\linewidth]{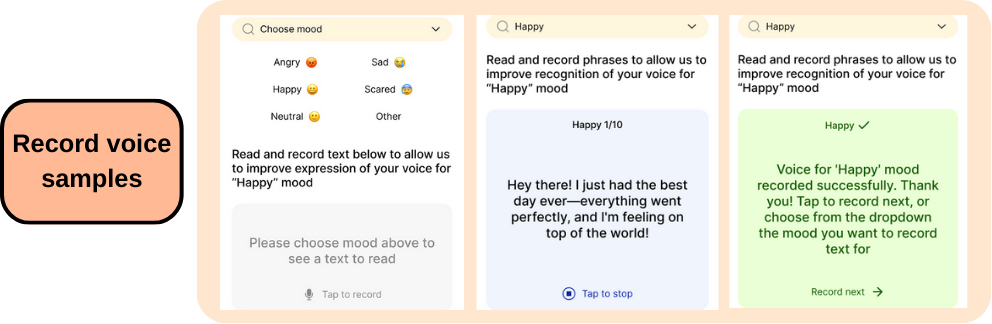}
    \caption{Voice Output Personalization Process}
    \label{fig:record}
\end{figure}

Personalization is a core aspect of \textit{Speak Ease}, ensuring that each user’s communication experience is tailored to their specific needs and preferences. To ensure personalization at the output stage, \textit{Speak Ease} offers personalized synthesized voice output that approximates users' natural voice. Users have the option to record samples of their own voice, which are then stored and used to produce emotionally expressive outputs that closely match their natural voice. Figure \ref{fig:record} shows the interfaces for the voice output personalization process, where users can select an emotion and record five sample sentences that reflect that specific emotional tone. Consequently, the speech generated by the TTS module mirrors the user's unique vocal characteristics, thereby enhancing the authenticity and emotional impact of their communication \cite{10.1145/3613904.3642080}. This approach ensures that each interaction maintains a personal touch, making the communication feel more genuine and personalized \cite{10.1145/3173574.3173857}.

Moreover, addressing the challenges of current AAC systems, which often produce robotic or impersonal outputs, the \textit{Speake Ease}'s outputs are designed to convey a wide range of emotional tones. This capability allows users to express their feelings more effectively and naturally, adding depth, authenticity, and realism to their communication \cite{10.1145/3613904.3642013}. 

\subsubsection{Technical details}
\begin{itemize}
\item \textbf{Voice sample storage:} Users record samples of their voice, which are stored in a secure Firebase database. These samples are used to train the TTS system to generate speech that mimics the user's natural voice. 

    \item \textbf{TTS system:} The system, powered by Eleven Labs \cite{ElevenLabs}, produces high-quality, expressive speech that matches the user's vocal characteristics. The TTS adjusts pitch, tone, and speed to convey the intended emotional state.
\end{itemize}

\subsubsection{Example scenario finished}
John has previously recorded several sentences expressing different emotions, such as happiness, during the app onboarding. For the message "I'm so happy to eat pizza tonight!", the TTS system uses the recorded samples to produce a voice output that is lively and matches the user's natural voice. The final output is an enthusiastic "I'm so happy to eat pizza tonight!" that sounds authentic and expressive.

In summary, our AI-integrated AAC system \textit{Speak Ease} aims to transform communication for users with atypical speech patterns by offering a comprehensive solution. It integrates multimodal input methods, including voice, touch, and keyboard, to cater to diverse user needs and enhance expressivity. By leveraging an advanced LLM, the system generates contextually relevant and emotionally resonant outputs. These outputs are delivered through TTS technology that replicates the user's natural voice, ensuring authenticity and emotional expressiveness. 

\section{Study: Evaluation with Speech and Language Pathologists}
The focus of this study was primarily on feasability study and evaluation of \textit{Speak Ease}'s intended functionality to support expressivity in AAC. \textit{Speak Ease} serves as a design probe to help us explore and understand how to better support and enhance users' expressivity when designing AAC systems. The study was IRB approved.

\subsection{Participants}
To better understand the role of expressivity in AAC systems and receive constructive feedback on \textit{Speak Ease}, one focus group (3 SLPs) and one interview (1 SLP) were conducted with subject matter experts - speech and language pathologists who had direct experience and expertise with AAC devices and work with speakers with impaired speech. \footnote{This arrangement was due to unavailability of 1 SLP to attend the focus group, yet their expertise and experiences brought significant value to our study} The study aimed to explore their perceptions of AAC expressivity, current system limitations, and potential improvements. Participants were guided through a structured discussion using a series of targeted questions, which addressed topics such as the definition and importance of expressivity, experiences with different input modalities, and the perceived value of voice customization.

\subsection{Procedure}
All participants were initially onboarded and asked to test \textit{Speak Ease} prior to their participation. During the focus group and follow-up interview, we gathered insights about their experiences using the application, with a particular focus on expressivity in AAC. We gathered participants’ feedback through remote video calls. We recorded each interview and took notes. Focus group sessions were conducted in a semi-structured format, which allowed for flexibility in discussion while ensuring key questions were addressed. Participants were invited to share their experiences and perspectives on expressivity in AAC communication, covering topics including input modalities, message delivery, and the use of personalized voice outputs. The sessions lasted approximately 60 to 90 minutes long and included open-ended discussions, demonstrations of AAC app functionalities, and follow-up questions to explore specific scenarios.

\subsection{Data Collection and Analysis}
Study sessions were video and audio recorded, with audio recordings transcribed using automatic speech recognition and manually verified for accuracy. These materials were then imported into NVivo, a qualitative data analysis software, to facilitate systematic organization of the data and coding. Transcripts and notes were coded to categorize participant feedback, focusing on key areas including input expressivity, multimodal inputs, and personalized voice generation. The findings provided actionable insights into how AAC devices could better support user expressivity.

\section{Findings}
%\section{Findings}

\subsection{Understandings of Expressivity in AAC}
Participants discussed the definition of expressivity in the AAC field and emphasized that expressivity in AAC involves communicating authentically through text, symbols, or speech generation. They also discussed the key factors that influence expressivity in AAC, including a user's level of literacy, nonverbal cues, and contextual conditions.

\subsubsection{Definition of Expressivity}
Participants defined expressivity in AAC as \textit{"being able to convey thoughts and feelings through writing or speech, and writing in the form of either text or symbols."} (P2). P1 further highlighted expressivity as \textit{"a very multimodal thing"}. P1 elaborated, \textit{"So the content of the message is there, but then how it's delivered, the facial expression, the tone of the voice, the body language, all of that, like the pitch or tension or whatever in somebody's voice can tell you a lot. So I think it's complicated."} Participants agreed that expressivity is not confined to the content of the message alone but also extends to the nuances of how the message is conveyed. P2 reinforced the complexity of expressivity, stating \textit{"(The needs for expressivity) depend on the person, and could be varied."} This variability is influenced by various factors as they described below. 

\subsubsection{Literacy as a Foundation for Expressivity}
Participants highlighted literacy as a key factor and the foundation for ensuring expressivity of AAC users. P3 mentioned,\textit{ "The only way that somebody is going to be able to say what's truly inside their head is if they can spell, if they can talk. So if somebody is literate and they can use text to speech or they can have word prediction."} Literacy allows AAC users to access AAC functionalities like text-to-speech and LLM word prediction, which enable more authentic expression. 

In contrast, illiterate users often rely on symbol-based systems, which inherently limit their expressivity. P3 highlighted the difficulties illiterate AAC users may face, stating \textit{"it's the vocabulary that they're restricted with. It's the actual words that the symbols represent or the thoughts and the concepts that the symbols represent (they cannot convey)."} P3 explained, \textit{"If they have to use symbols because they're not literate, then they're only going to be able to use the symbols that somebody else gave to them, that somebody else put in their system. So then their vocabulary will be limited."} For users who cannot achieve literacy, collaboration with their close communication partner is vital to ensure that symbol-based vocabularies in AAC systems reflect their needs and interests as closely as possible. P3 added, \textit{"So that's why if somebody's not literate, we do our very best to get vocabulary ideas from the people who live with that person and who know that person really really well, so that we can attempt to give them the best vocabulary that will meet their needs and their interests. But as long as somebody cannot spell, they're always going to be at the mercy of the vocabulary that someone else gives them."} 

Besides literacy, nonverbal cues such as facial expression, body language, and voice tone are identified by participants as key components for users to be expressive.

\subsubsection{Contextual Factors that Affect AAC Expressivity}
Participants emphasized how variables such as communication partners and environmental conditions could greatly influence how effectively AAC users express themselves. These contextual factors can be broadly categorized into two types: the AAC output interpretation by the communication partner, and the physical and emotional states of the AAC user.

\textit{The Interpretation of the Output by Communication Partners.} Human interpretation of the AAC output plays a key role in ensuring AAC users' expressivity. Participants noted that certain users are only fully understood by specific individuals, such as close family members, but struggle to communicate effectively with others in the community. Specifically, P2 and P3 both agreed that \textit{"if you have someone living in a family, they would develop their own system of how they express within that context of their close family"}. P3 also stated that \textit{"some people (AAC users) might only be understood by their significant other or their parent or their child, but nobody else. The minute they leave the house and go out into the community, nobody understands them."} This stresses the necessity for AAC systems to include features that allow users to select their intended speaking partner and adapt various communication approaches accordingly, enabling them to be better understood outside their close social circles, which ultimately could enhance users' expressivity under different scenarios.  

\textit{Users' Physical and Emotional Factors.} Participants also identified factors like fatigue and stress as barriers to effective AAC communication. Users might struggle to express themselves clearly as the day progresses or under certain emotional states. These factors vary by individual and situation. P2 stated, \textit{"They (patients) all have complex communication needs and require some support with either comprehension expression or both. But it could vary depending on the individual. They could have no ability to speak verbally or it could be dependent upon fatigue or time of day or environment."} P4 also elaborated, \textit{"Fatigue can reduce the volume or (clarity of) speech articulation, because if you're tired and you already have impaired articulation, then it might weaken the articulation even more. So it could become more imprecise, more clumsy."} P4 added, \textit{"all of my clients with acquired conditions (that caused speech impairments) talk about the variability, fatigue, medication, mood."} The findings emphasized the need for AAC systems to adapt to users’ changing circumstances and fluctuating states, providing flexibility in expressivity based on various environments and conditions at any given time.

\subsection{Tradeoffs and Challenges in Enhancing Expressivity of Speak Ease Input Methods}
\subsubsection{Benefits and Challenges of Multi-modalities Input in AAC}
Participants agreed that the option of having access to and enabling switching between different input modalities from speech, text, and emojis would be beneficial to users. This flexibility is particularly vital because AAC users' speech capabilities often change gradually, they might need to switch input modes based on their symptoms. Participants also pointed out that the multi-modalities input is valuable especially for "savvy users" who have experience using AAC apps. As for those who are not familiar with AAC systems or tech devices in general, education and guidance are necessary. P2 emphasized the importance of consistent support for these non-savvy users, stressed \textit{"The barrier (for non-savvy users) is whether they have supporters who can be consistent with teaching and implementation. And it would depend too on what their history has been in terms of having tried systems in the past and a whole bunch of factors."} The challenges were consistently brought up by participants when discussing the implementation and actual usage of different input modes and functionalities as follows.

\subsubsection{Speech Input: Need to tailor for dysarthric speech}
Participants emphasized the importance of \textit{Speak Ease}'s ability to process and understand dysarthric speech as input. P3 stated, \textit{"From the speech input point of view, it would depend on how well the app listens to speech that we would say is not normal or dysarthric}." Participants emphasized that this is particularly critical for users with degenerative conditions like ALS or cerebral palsy, where their speech quality deteriorates over time.

\subsubsection{Emoji Input: Enhancing Expressiveness but Remaining Supplementary}
Emojis offer an alternative way for users to communicate using \textit{Speak Ease}. However, emojis were often seen as supplementary to core message delivery rather than a primary input mode. Participants noted that while emojis might add richness to a conversation, the primary focus for many AAC users remains on "getting the messages out". For some AAC users, especially those with cognitive or developmental disabilities, selecting the right emoji might not be intuitive. As a result, emojis may become a secondary option in comparison to more straightforward text inputs. P3 claimed that \textit{ "emoji is probably the least valuable for a face-to-face conversation... It’s not going to be very usable to a large group of people."} and added, \textit{"If they have to use symbols because they're not literate, then they're only going to be able to use the symbols that somebody else gave to them that somebody else put in their system. So then their vocabulary will be limited."} Besides, emojis have their own limitations in conveying meanings. P4 mentioned, \textit{"Emojis are a really impoverished version of (symbolic communication in AAC) because there're no real verbs. It doesn't represent all the group categories of language, but it has some.}" Therefore, while emoji input can potentially contribute to the expressiveness of AAC communication, its actual usage depends on individual preferences, cognitive abilities, and the specific communication context. 

\subsection{Opportunities and Challenges in Enhancing Expressivity with LLMs in Speak Ease}
\subsubsection{Adaptive Learning vs. Preprogrammed Systems}
Participants discussed whether the system should rely on preprogrammed inputs or adaptive AI learning in \textit{Speak Ease}. Preprogrammed configurations were seen as impractical. P2 explained, \textit{"If it involves us programming it all in, then there are other options we would choose based on the amount of clinical time it would require."} Conversely, adaptive AI systems like LLMs, which are capable of learning from user input over time, were identified as a more feasible solution. However, P4 raised questions about whether LLM is "smart enough to learn", adding \textit{"I think often when there are errors, they're pretty inconsistent."} This shows that while adaptive LLM learning is promising, it also faces challenges. The system must be capable of distinguishing between useful patterns and noise in user input.

\subsubsection{LLM Word Suggestion: Reducing effort but not universally beneficial} 
The word prediction and suggestion feature enabled by LLM in \textit{Speak Ease} is intended to facilitate AAC communications. Users are able to enhance typing efficiency and produce clear and contextually accurate output that is easily understandable by others. However, participants noted that the feature may not be universally beneficial. Users are likely to forgo this function for the sake of efficiency. P4 claimed, \textit{"I think having it suggest vocabulary and then having to read the suggestions and find the one they want. Sometimes it's just faster just to type it."} Similarly, P2 also commented, \textit{"It depends on access too and how efficient they are. So having prediction is helpful to increase efficiency when you're typing for sure."} P3 added, \textit{"If somebody is literate and they can use text to speech, (then) they can have word prediction."} Participants also stressed to this date there are no effective ways for LLM to interpret input errors. P2 added, \textit{"If you don't get the first two letters correct, it is not going to pick the right word. You're not going to get that as an option. So that's the issue that we see in terms of spelling and use of prediction. You have to be able to get those first two."} This highlights how the effectiveness of word prediction by LLM depends on the time frame of the communication setting, as well as individual user preferences, abilities, and familiarity with such features. This highlights how the effectiveness of word prediction by LLM depends on the time frame of the communication setting, as well as individual user preferences, abilities, and familiarity with such features.

\subsubsection{Generate Context-aware Suggestions: Requiring High Meta-awareness}
LLM generates contextually appropriate sentences based on user-specified communication partners and intended emotions in \textit{Speak Ease}. However, participants noted that the feature may require advanced metacognition or meta-awareness from users to be fully effective. P2 commented on the high cognitive demands of this feature, \textit{"This is going to need a very high-level user to be able to differentiate and then go in and choose these things. So a lot of forward thinking, I'm communicating with someone new, I want it to sound a certain way." } Participants also noted that this functionality might be more beneficial for asynchronous communication settings, such as composing a message for a doctor or teacher, where users have more time to select appropriate tones and refine expressions.

\subsubsection{Contextual Emotional Expression: Struggles in Conveying Emotional Expressions Accurately}
Participants expressed interest in the ability to select emotions to attach to a given message in \textit{Speak Ease}, recognizing the potential for LLM to add emotional nuances to enhance expressivity. P2 appreciated the feature and stated, \textit{“I quite liked that you could choose the emotion that would be attached to a given message.”} However, P4 questioned how effectively the emotions were conveyed through the process: \textit{"The (selected) emotion didn't come through that much. Does the message reflect the emotion I want to convey? I felt sort of more so for happy and neutral, but less so for sad and angry."} Similarly, P3 added, \textit{"I found that the emojis didn't change my voice much at all. So I didn't hear a big difference between them."} This reveals a gap between our intention of emotional expression features and the practical execution. Challenges were raised regarding the accuracy and appropriateness of different emotional expressions through the LLM process. Participants questioned whether the AAC system would consistently and accurately interpret and reflect users' designated emotions, especially for users with varying emotional recognition abilities or cognitive limitations.

\subsubsection{Concerns about Facilitated Communication (FC) in AI-powered AAC systems}
Participants raised serious concerns regarding the ethical implications of AI-assisted communication in AAC, particularly in the context of "facilitated communication". Facilitated Communication (FC) is a technique whereby individuals with disabilities and communication impairments allegedly select letters by typing on a keyboard while receiving physical support, emotional encouragement, and other communication supports from facilitators \cite{schlosser2014facilitated}. The validity of FC stands or falls on the question of who is authoring the typed messages, whether the individual with a disability or the facilitator \cite{schlosser2014facilitated}. This raised the question of ownership and authenticity in AI-generated content using AAC tools. The group also referenced other AI-powered AAC innovations that raised ethical concerns, such as systems that generate contextual suggestions based on environmental cues. P3 mentioned, \textit{"(There is) the app that actually knows where you are in a room or if you're in a restaurant or if you're in a medical environment, and will generate options to choose from based on where you are in an environment. And everybody said this sounds like facilitated communication. Who is the author at the end of the day?"} 

Participants highlighted that authentic communication must originate from users themselves, especially in serious conversations or under sensitive circumstances. P2 stated,\textit{ "Be careful of conversations that are far more serious outside of small talk where authenticity and ensuring that the message is actually coming from the user."} P3 continued to explain, \textit{"The impression of leading the conversation from the AI would be in question in a court of law as an example."} They recounted an example of a client who sought AAC assistance for "targeted vocabulary" related to a legal incident.

The risks of blurring lines between AI assistance and user-autonomous communication led participants to suggest strict boundaries. P4 mentioned content filtering mechanisms could potentially be useful to block vocabulary related to inappropriate or sensitive topics. \textit{"This is not a fully developed thought, but just in terms of AI potentially leading to a situation where there's facilitated communication. I guess we filter out certain content in a way... with the individual who's asking for support, for vocabulary to support a situation of abuse. We're not going to do that because we're not going to lead the conversation. So I think we automatically do that."}

\subsection{Enhancing Expressivity by Improving Personalized Voice Output in Speak Ease}
\subsubsection{Personalized Voice Output: Natural Sounding but Challenging}
Several participants spoke highly of the voice banking feature of \textit{Speak Ease} to be "really good". The feature allows users to create voice outputs that resemble their natural voice. P2 mentioned the positive experience with this feature, \textit{"It did sound so much like me, so I could see myself using that in some situations for sure."} P4 also remarked on the resemblance of the voice output: "\textit{I just liked that it sounded like me...I think that might be me. And then I heard another one after in a different emotion that sounded very much like me.}" P1 added, \textit{"it's not something that has been raised a lot to me, but people are happy in some cases if they can use banked voice so that it sounds like them."}

However, participants also noted challenges in generating emotion-specific voice outputs. P1 stated, \textit{"I think there were a few sad that didn't sound as expressive as the happy, but then I don't know if it's based on my recording and it is a little awkward. I was like, how do I record a sad? So maybe some support for that for people in the future too."} This underscores the need for user guidance during the voice banking process, especially in capturing diverse emotional expressions effectively. Alternatively, the unnatural and awkward experience users may face during the voice banking process could be alleviated by using preexisting recordings from their past. 

Furthermore, P1 emphasized the diverse individual choices regarding voice banking. While some users find it crucial to retain their unique voice identity, others may not prioritize it, particularly depending on their personal circumstances or stage in their diagnosis. As P1 explained, \textit{"That (voice banking) is a really individual thing. I would say probably less than half, maybe a third of the people that I offer it to are actually interested in it. For some people... it doesn’t matter what my voice will sound like if I need a device in the future. And for some people, it’s really, really important."} The reasons for these varied responses were linked to emotional readiness and acceptance of future changes in communication abilities. P1 elaborated, \textit{"It might just also be because they’re not in a place maybe in their diagnosis where they’re accepting what might happen in the future with their communication... it does seem that for some people it doesn’t matter, and for some people, it really does, but it goes both ways."} P4 also added, \textit{"Voice banking is a sensitive topic because it's associated with a huge loss. So honestly, some people don't even want to talk about it. Some people break down in tears, they don't want to hear about it. Other people are like, yeah, I want to bank my voice right away. Again, it's a very personal thing."}

Therefore, while the feature of generating personalized voice output could potentially enhance expressivity, its actual impact relies heavily on how well it aligns with the personal preferences and emotional circumstances of each user.  

\subsubsection{Exploring Alternative Voice Capture Methods}
Participants explored alternative methods for voice capture instead of voice banking, which is to use pre-recorded voices or audios for generating personalized voice output. P3 noted the potential for using existing voice recordings from public figures, such as those found in movies, television, or online lectures. \textit{"Some of the voice recording apps will take voice recordings from people. So people in the public eye as an example, people that have recordings on television or have been in movies or any of that sort of stuff, they can take those recordings and make voices out of them. Teachers, lecturers, profs that have had their lectures recorded, especially now that a lot of coursework is online. So if the online courses have been recorded and kept, those sorts of things can be used to create voices, but it's really past recordings that people would be relying on."}

The discussion expanded to consider the potential for augmenting and improving the intelligibility of the voice of individuals with developmental disabilities. P1 noted, \textit{In the case of somebody with a developmental disability, it's no longer preservation because they have what they have. It would be more like augmentation, like improvement."} Augmenting voice focuses on improving the clarity and expression of speech for individuals who may already have speech impairments. P1 referenced a TED talk about a group in the United States capable of taking minimal vocalizations (such as "ah" or other basic sounds) from individuals with no clear articulation and creating a more intelligible voice through the use of donor voices. As P1 explained, \textit{"It's very different (from voice banking), because voice banking is to preserve what you're going to lose. So it's a totally different thing."}

\section{DISCUSSION}
%\section{Discussions}

This study looked into how multimodal input approaches integrated with LLMs can enhance expressivity in AAC systems. The findings above demonstrate the challenges and opportunities in improving users' expressivity in AAC systems. Multiple factors must be taken into consideration during the design process. In this work, we reflect on the ways in which expressivity in AAC could be better supported at the input and output stages.

\subsection{Recognizing the Role of Communication Partners}
Building on our study findings, we realize the importance of recognizing communication involves not just the individual using the AAC system, but also their communication partners. Fer et al. \cite{fer2018verbal} defined communication as the two-way process of reaching mutual understanding, in which participants not only exchange (encode-decode) information, news, ideas and feelings but also create and share meaning. Previous AAC system designs mainly focus on catering needs of AAC users with speech impairments \cite{curtis_state_2022}. However, we discovered that designing effective AAC tools also requires careful consideration of the roles played by all communicators and the contexts in which these interactions occur. As our study participants highlighted, communication partners often play a key role in educating, facilitating, and interpreting messages generated through AAC systems. These partners may include close family members, caregivers, friends, educators, or strangers in public settings. Their engagement and understanding significantly impact the effectiveness of AAC systems. Therefore, when designing AAC systems, it’s essential to acknowledge these roles and create tools that not only empower the user but also support their communication partners. Providing resources, tutorials, or in-app guidance for partners can enhance their understanding and foster more meaningful interactions. Moreover, communication partners should have input in customizing the AAC system such as creating commonly used phrases or symbols to contribute to smoother communication. By supporting both the AAC user and their communication partners in a holistic approach, AAC systems could serve as a bridge between users and the outside world, fostering collaboration, mutual understanding, and natural interactions.

\subsection{Balancing Accessibility and Complexity when Designing Multimodal Input to Enhance Expressivity}
Our AAC app offers users the option to input through voice, text, or emojis. Participants universally agreed that the availability to choose different input modalities is valuable for enhancing user expressivity, especially for experienced users who are familiar with the system. However, challenges were also noted in implementing multimodal input systems, especially for those users who are not as tech-savvy. While multimodal input options increase accessibility, they can also introduce complexity, requiring users to engage with multiple interfaces or modes of input. This can be overwhelming for some users, especially those with cognitive disabilities, who may struggle or feel overwhelmed to manage different input methods \cite{10.1145/2998181.2998284}. Therefore, users may choose not to access the AAC system if they have experienced difficulties of using the system in the past, which could ultimately harm their expressivity. To address this issue, ensuring proper guidance and education on the use of AAC is important. Participants pointed out that the barrier is whether AAC users have supporters who can be consistent with teaching and implementation. 

Additionally, the choice of input method is also influenced by the AAC user's diagnosis and their condition's progression. People with limited speech or in the later stages of a progressive condition may exhibit very different communication patterns and needs \cite{10.1145/2998181.2998284}. For example, someone with aphasia or cognitive disabilities might rely more on text-to-speech features instead of voice input. Thus, while multi-modality input can potentially enhance expressivity and efficiency, its effectiveness depends on individual user needs, capabilities, and the ability of the system to integrate inputs smoothly. The challenge lies in creating systems that balance flexibility with ease of use, ensuring that all users can take advantage of the benefits without being overwhelmed by the complexity of multiple input options.

\subsection{Balancing Efficiency and Personal Needs with Opportunities for Expressivity}
Efficiency remains to be a key consideration in AAC, with users often opting for the fastest and easiest method of communication. Participants claimed that while ensuring expressivity is important, efficiency and time constraints are the top priorities for many AAC users. For example, instead of setting up contextual factors or selecting stored phrases, users may choose to directly type to convey their message quickly. The challenge remains in balancing the need for efficient communication with the desire for more expressivity. 

The urgency of real-time interactions and fatigue further impact expressivity. AAC users often have to decide how much they want to say and consider both the amount of time they will require before they decide if they want to compose the message \cite{10.1145/2998181.2998284,valencia_2023}. While some users appreciate features like speech synthesis that sounds like their natural voice, they often forgo these features to reduce the time and effort spent on message delivery. The need for speed and ease of communication tends to outweigh the desire for expressivity in many use cases, as users prioritize simply getting their message across as quickly as possible. Therefore, when designing features in AAC systems, we need to balance efficiency with opportunities for expressivity.

Alternatively, we could explore the possibility of users using AAC systems for asynchronous communication, such as drafting an email or preparing a script, where they have more time to craft the messages ahead. In asynchronous settings, users would have more opportunities to use functionalities in AAC systems such as voice customization and context-specific phrasing to enhance expressivity without the pressures of immediacy.

Overall, there is no one-size-fits-all solution when designing the AAC system. The effectiveness of the AAC app needs to be adapted to meet the diverse needs of different users. In particular, expressivity in AAC should be tailored based on factors such as the user's diagnosis, stage of the condition, cognitive abilities, mental and physical states, and personal preferences. It would also depend on the real-time or asynchronous user cases. Therefore, maintaining flexibility and adaptability in the design of AAC systems is essential to support effective and personalized expressivity.

\subsection{Authenticy and Ethical Implications of Using AI in AAC}
While leveraging LLM could help AAC users communicate more efficiently, our participants raised concerns about the role of AI during the process regarding the impact on the user's authenticity and agency. Previous studies have looked into the subject. \cite{Cardon_2023} claimed that AI-assisted writing may hinder individual uniqueness and authenticity in aspects like voice, style, and tone. \cite{kadoma_2024} explored the influence of writer-style bias on inclusion, control, and ownership when collaborating with LLMs. Choosing a generated phrase made AAC users believe they did not have agency over their own choice, rather the system had made the choice for them. Agency is a critical component of allowing AAC users to fully express themselves \cite{Vertanen_2015}.

Therefore, we need to be cautious about AI's roles and responsibilities when designing AAC systems. As our findings suggested, LLMs could potentially lead the conversation or dictate the messages, which could be legally and ethically problematic in sensitive situations, such as legal or academic settings. There are also challenges regarding AI misinterpreting users' intent, such as mistranslating speech or generating unwanted words, which could lead to embarrassment or miscommunication. It is imperative to ensure AAC users retain full control over their communication and the content generated, while also avoiding situations where AI could lead or manipulate users' messages. The distinction between providing supportive stored messages and leading conversations remains critical. To achieve this, Kadoma et al. \cite{kadoma_2024, Weinberg_2024} revealed that it is crucial to increase tech awareness and literacy of what AI tools are capable of since a majority of people are still not aware of what AI can or cannot do. 

While AI holds great promise in the AAC field, it must be operated within clear boundaries and implemented thoughtfully to avoid ethical issues. Further study needs to focus on adopting methods to ensure AI does not dominate the communication process so that user autonomy would not be compromised. 

\section{LIMITATIONS AND FUTURE WORK}
While our prototype \textit{Speak Ease} demonstrates the feasibility of integrating expressivity into AAC systems, we acknowledge that \textit{Speak Ease} is not yet a fully developed application intended for deployment. We leveraged this early-stage system as a design probe to investigate the priorities and needs of enhancing expressivity for AAC users. Future work needs to focus on iteratively refining the system based on feedback and incorporating advanced features. 

To better evaluate and get feedback for \textit{Speak Ease}, it would be beneficial to recruit a diverse group of users with various diagnoses to test \textit{Speak Ease}. Potential users could include individuals with Down syndrome, Parkinson’s disease, and rapidly progressing conditions like ALS, which present unique challenges such as low volume, monotone pitch, and rapidly changing communication needs.

Moreover, a key area for further development could be made to enhance the emotional expression feature in \textit{Speak Ease} to ensure that selected emotions by users are accurately conveyed and perceived by listeners. This could involve refining voice synthesis or incorporating more dynamic visual cues.To expand the user cases of \textit{Speak Ease}, future work could continue exploring suitable settings to support users in asynchronous communications. 

Furthermore, future work could keep exploring how to systematically use existing recordings to generate personalized voice outputs. This includes developing methods for extracting and synthesizing voices from existing video or audio sources while ensuring ethical considerations, such as obtaining permissions and addressing privacy concerns. 

Another important direction of future work is to explore how to enhance the contextualization of LLM responses within the AAC system. As users interact more frequently with \textit{Speak Ease}, the model could potentially store previous conversations and adapt responses based on the user’s specific context and interaction history. Conversations with different individuals will be retained, enabling the model to generate contextually appropriate responses that align with the user's past communication patterns. This customization aims to improve the relevance and accuracy of follow-up responses by factoring in prior interactions.

\section{CONCLUSION}
In this paper, we presented \textit{Speak Ease}, an augmentative and alternative communication (AAC) system to support users' expressivity by integrating multimodal input, including text, voice, and contextual cues (conversational partner, and emotional tone), with the large language models (LLMs). Through feasibility study with 4 SLPs we highlighted the importance of ensuring expressivity in AAC systems. Our findings present the potential of using multiple input modalities and LLM to support expressivity. However, we also identified the directions for additional tailoring of presented AAC system to further cater to individual needs, abilities, and contexts of AAC users. In particular, we identified the need for a careful  balancing between technological complexity and user's capabilities. %Moving forward, our work will focus on refining \textit{Speak Ease} by addressing the limitations and opportunities outlined above. 

\bibliographystyle{IEEEtran}
\bibliography{reference}

\end{document}